\documentclass[final,3p,sort&compress]{elsarticle}

\usepackage{graphicx}

\usepackage{amsmath,amssymb,amsfonts}
\usepackage{bm}

\def\be{\begin{equation}}
\def\ee{\end{equation}}
\def\ba{\begin{eqnarray}}
\def\ea{\end{eqnarray}}
\def\bd{\begin{displaymath}}
\def\ed{\end{displaymath}}
\def\bq{\begin{eqnarray}}
\def\eq{\end{eqnarray}}

\journal{Surface Science}

\begin{document}

\begin{frontmatter}

\title{Adsorbate surface diffusion: The role of incoherent
tunneling\\ in light particle motion}

\author{A. S. Sanz$^{1,2}$\corref{corresp}} \ead{asanz@iff.csic.es}

\author{R. Mart\'{\i}nez-Casado}

\author{S. Miret-Art\'es$^1$}

\cortext[corresp]{Corresponding author}

\address{$^1$Instituto de F\'{\i}sica Fundamental (IFF--CSIC),
Serrano 123, 28006 Madrid, Spain}

\address{$^2$Department of Physics and Astronomy, University College
London, Gower Street, London WC1E 6BT, United Kingdom}

\address{Thomas Young Centre, Department of Chemistry,
Imperial College London, South Kensington, London SW7 2AZ, UK}

\begin{abstract}
The role of incoherent tunneling in the diffusion of light atoms on
surfaces is investigated. With this purpose, a Chudley-Elliot master
equation constrained to nearest neighbors is considered within the
Grabert-Weiss approach to quantum diffusion in periodic lattices.
This model is applied to recent measurements of atomic H and D on
Pt(111), rendering friction coefficients that are in the range of
those available in the literature for other species of adsorbates. A
simple extension of the model has also been considered to evaluate
the relationship between coverage and tunneling, and therefore the
feasibility of the approach. An increase of the tunneling rate has
been observed as the surface coverage decreases.
\end{abstract}



\end{frontmatter}



\section{Introduction}
\label{sec1}

Hydrogen diffusion on metal surfaces is being the subject of an
intensive applied and fundamental research for many years
\cite{graham,lauhon,wang,sundell,zheng,chado,jardine}. On the one
hand, the involvement of this process in the design of fuel cells
or in the storage of H$_2$ makes it of technological interest. On
the other hand, at a fundamental level, a good understanding of
this dynamics is essential, for it provides important information
on the adiabatic interaction between adsorbate and substrate.
Indeed, because atomic hydrogen is the simplest element that may
undergo chemisorption, its motion on a weakly corrugated surface
constitutes a benchmark to study two possible competing diffusion
processes: classical (activated) {\it over-barrier} hopping {\it
versus} quantum-mechanical {\it under-barrier} tunneling. These
two diffusion mechanisms are intimately related to the role played
by phonons and conduction electrons. In the classical regime, the
rate decreases with the inverse of the temperature until reaching
the quantum/tunneling regime, where the corresponding rate
approaches a constant, temperature-independent value. Apart from
bulk or surface diffusion problems, this behavior is also typical
in many different chemical reactions or processes
\cite{goldanskii,hanggi,weiss}.

Motivated by recent measurements of atomic hydrogen on a Pt(111)
surface \cite{jardine}, here we tackle this surface diffusion
problem assuming that the process is essentially ruled by deep
tunneling. The adsorbate quantum motion between neighboring sites
will take place mainly through incoherent tunneling for not too low
surface temperatures. As shown elsewhere \cite{chudley,salva}, in
this regime the Chudley-Elliott model results very convenient to
describe the corresponding (surface) dynamics. Now, in order to
include properly the effects of deep tunneling and temperature, we
have also made use of the approach developed by Grabert and Weiss to
account for quantum diffusion in periodic potentials
\cite{weiss,grabert1,grabert2}. In particular, the so-called bounce
technique considered by these authors has allowed us to establish
the connection between this approach and the Chudley-Elliott model,
and therefore to obtain an analytical expression to account for
transition rates in conditions of deep tunneling. The application of
this model to the aforementioned experimental data has rendered a
good agreement between theory and experiment in a fitting procedure.
More specifically, we have been able to extract friction
coefficients for the two adsorbed species considered in the
experiment, namely H and D. The values obtained are in consonance
with the isotopic effect ($\gamma_{\rm D}$ is about a factor two
smaller than $\gamma_{\rm H}$) as well as with those reported in the
literature for other species \cite{jardine2}, which are of the order
of picoseconds. Furthermore, in order to investigate the effects of
the surface coverage on tunneling diffusion, and corroborate the
friction coefficients extracted from the fitting procedure, we have
also carried out an alternative analysis by means of a
temperature-dependent, collisional-friction model
\cite{ruth1,ruth2,ruth3}, shown to be valid at low coverages (as it
is the case of the experiment, where $\theta = 0.1$~ML, and below).
In this model (the so-called two bath model), the adsorbate undergoes
a total friction resulting from the sum of two contributions, namely
the usual substrate friction plus a collisional friction accounting for
the collision among adsorbates.
From its application, we have observed that
tunneling rates increase as the coverage decreases, which would be
consistent with the intuitive fact that tunneling between
neighboring sites becomes faster and, therefore, more relevant.


\section{Theory}
\label{sec2}

Diffusive surface dynamics is usually described in terms of a series
of discrete jumps undergone by the adparticle when it moves on a
two-dimensional periodic lattice of binding sites. In the case of
activated diffusion, when the thermal energy is higher than the
barrier height separating neighboring sites, adparticles are mainly
assumed to perform discrete over-barrier jumps between sites. In the
specific case of the Chudley-Elliott model, this diffusion dynamics
is accounted for by a master equation in terms of the van Hove
$G({\bf R},t)$-function or time-dependent pair correlation function
\cite{vanHove}. This function is widely used to describe statistical
ensembles of interacting particles, thus generalizing the well-known
pair distribution function $g({\bf R})$ from statistical mechanics
\cite{mcquarrie} by providing information about the interacting
particle dynamics. In other words, given a particle at the origin at
some arbitrary initial time $t = 0$, $G({\bf R},t)$ gives the
averaged probability of finding the same or another particle at the
surface position ${\bf R}$ at time $t$.

The same approach can be extended to analyze under-barrier,
tunneling transmission. This is the behavior expected for lighter
particles, such as atomic hydrogen, and thermal energies lower than
the barrier height, where tunneling-mediated diffusion is assumed to
be dominant in the dynamics between nearest neighbors
\cite{jardine}. In such a case, incoherent tunneling can be
described by means of a general master equation of the form
\begin{equation}
 \dot{G}({\bf R},t) = \sum_{\bf j} \nu_{\bf j}\ G({\bf R}+{\bf j},t) ,
 \label{G}
\end{equation}
where $\nu_{\bf j}$ accounts for the tunneling rate involved in the
transition between the lattice point ${\bf R}$ and the nearby one
${\bf R}+{\bf j}$, with ${\bf j}$ being the jump vector among
different lattice points running over all lattice vectors (positive,
negative, and even zero).

The space Fourier transform of the $G$-function is the intermediate
scattering function,
\be
 I(\Delta {\bf K},t) = \langle e^{-i \Delta {\bf K} \cdot {\bf R}(t)}
    e^{i \Delta {\bf K} \cdot {\bf R}(0)}\rangle ,
 \label{eq:DSF}
\ee
with the brackets denoting an ensemble average. This function
measures the time correlation loss for a given parallel (along the
surface) momentum transfer of the probe particle, $\Delta {\bf K}$.
Therefore, it can also provide us with information about friction
coefficients at different coverages and (surface) temperatures along
with the observed $\Delta {\bf K}$ direction. Without loss of
generality, we can assume the diffusion process among the different
wells formed by the surface corrugation is one-dimensional along
this direction. Hence the intermediate scattering function can be
expressed as a Fourier series, as
\begin{equation}
 I(\Delta K_\| ,t) = \sum_n G_n (t) e^{i \Delta K_\| n} ,
 \label{I}
\end{equation}
where $\Delta K_\|$ is a dimensionless momentum transfer resulting
from the projection of the lattice vector ${\bf j}$ along the
direction pointed by $\Delta {\bf K}$ multiplied by the lattice
constant $a$, i.e., $\Delta K_\| = a \|\Delta {\bf K} \| \cos
\alpha$, with $\alpha$ being the angle between $\Delta {\bf K}$ and
${\bf j}$. Only first neighbors are considered. Therefore, in the
particular case of the Pt(111) lattice geometry \cite{graham}, for
four of these neighbors $|\Delta K_\| | = a \Delta K \cos (\pi/6)$,
while for the other two, $\Delta K_\| = 0$. Regarding $n$, it labels
the $n$-th well of the binding site (bearing in mind this
tight-binding like model).

Taking this into account, for nearest neighbors, Eq.~(\ref{G}) can
be expressed in terms of the $G_n(t)$ coefficients, as
\be
 \dot G_n (t) = \nu^+_{n-1} G_{n-1} (t) + \nu^-_{n+1} G_{n+1} (t)
  - (\nu^+_n + \nu^-_n ) G_{n} (t) .
 \label{diffG}
\ee
Here, $\nu^\pm_{n \mp 1}$ denotes the tunneling transition rate from
the $(n \mp 1)$-th well to the $n$-th one, while $\pm$ account for
the tunneling rates to the right or left neighboring well,
respectively. Tunneling rates are assumed to be equal for the left
or right direction and independent of the well site. The
differential equation (\ref{diffG}) can be solved analytically for
the initial conditions $G_n (0) = \delta_{n,0}$. From this solution,
we find that the intermediate scattering function can be recast as
\be
 I (\Delta K_\|,t) = e^{- 2 \bar{\Gamma} t \sin^2(\Delta K_\|/2)}
  = e^{- \bar{\Gamma} t} e^{\bar{\Gamma} t \cos(\Delta K_\|)}
  = e^{- \bar{\Gamma}t} \sum_{n = - \infty}^\infty
    I_n(\bar{\Gamma}t) e^{i \Delta K_\| n} ,
 \label{ISF}
\ee
where $I_n$ is the modified Bessel function of integer order $n$ and
$\bar{\Gamma} = \nu/2$ describes the global tunneling rate. This
standard form is in agreement with the models found in the
literature to describe surface diffusion.

Equation~(\ref{diffG}) is the same equation already obtained by
Grabert and Weiss \cite{weiss,grabert1,grabert2} to describe quantum
diffusion in periodic lattices. By means of the so-called bounce
technique, they found an analytical expression for the tunneling
transition rate between adjacent sites, which reads as
\be
 \bar \Gamma = \frac{\sqrt{\pi}}{2} \frac{\Delta^2}{\omega_0}
  \left( \frac{\pi k_B T}{\hbar \omega_0} \right)^{2 \zeta - 1}
  \frac{\Gamma (\zeta)}{\Gamma (\zeta + 1/2)} ,
 \label{fit-function}
\ee
where $\Gamma(\zeta)$ denotes the Gamma function of the
dimensionless friction coefficient,
\begin{equation}
 \zeta = \frac{m a^2}{2\pi\hbar}\ \! \gamma ,
 \label{zeta}
\end{equation}
$\omega_0 = V''(r_{\rm min})/m$ is the well harmonic frequency, $m$
is the adparticle mass, and $\Delta$ is the dressed tunnel matrix,
which is a function of the so-called bare tunnel matrix
\cite{weiss}, $\Delta_0$. Computing the diffusion coefficient
through the simple relation
\begin{equation}
 D = a^2 \bar{\Gamma} ,
\end{equation}
one clearly sees that the usual Einstein relation does not hold
anymore in the case of tunneling, since it does not scale with the
friction coefficient as $\gamma^{-1}$.

The characteristic temperature power law for tunneling rates given
by Eq.~(\ref{fit-function}) was first proposed by Grabert and Weiss
\cite{grabert1,grabert2} and Kondo \cite{kondo} when analyzing the
non-adiabatic response of the conduction electrons. This relation is
claimed to be valid at any temperature (including zero temperature)
provided $\zeta > 1$. This fact is precisely what warranties the
applicability of the Grabert-Weiss approach here. As it will be seen
in the next Section, the value of $\zeta$ obtained for both
adsorbates is such that this relation effectively holds, and
therefore that this approach is valid to describe the experiment
along the whole range of temperatures considered in it.

According to the diffusive model here considered, the loss mechanism
may come from the lattice relaxation and/or the electronic
contribution. Hence, the spectral densities as well as the
corresponding friction coefficients are additive. In principle, the
power law (\ref{fit-function}) is a general result for Ohmic
friction regardless of the loss mechanism and for surface
temperatures below the crossover temperature. As reported by Sundell
and Wahnstr\"om \cite{sundell}, this typical power law was first
suggested by Kondo \cite{kondo} and seems to be related to the
electronic contribution, i.e., the non-adiabatic response of the
conduction electrons to the adparticle (H or D) motion. However, a
similar behavior can also be found when the lattice relaxation is
replaced by a non-Ohmic dissipation \cite{grabert3}. In view of the
temperature-dependent behavior discussed by Grabert and Weiss
\cite{grabert1,grabert2}, and also presented here, the use of this
power law seems to be quite more general and valid for any loss
mechanism or the addition of several uncorrelated loss mechanisms
(which implies the sum of friction coefficients due to electrons,
phonons, and adsorbates) than Wolynes' dissipative transition state
theory \cite{hanggi,wolynes}.


\section{Results}
\label{sec3}

In order to test the feasibility of the above approach and therefore
the role of tunneling in adsorbate surface diffusion, we are going
to analyze the experimental data obtained \cite{jardine} for atomic
H and D on Pt(111) at a low coverage ($\theta = 0.1$~ML) and a range
of temperatures between 80~K and 250~K. The observable in these
experiments is the so-called polarization function, which is
essentially the intermediate scattering function (\ref{ISF}) (or, at
least, proportional to it). Thus, by means of a simple fitting of
the experimental data to a functional form $I(t) = A e^{-\alpha t} +
B$ for Eq.~(\ref{ISF}), for a fixed coverage, momentum transfer, and
temperature, the dephasing rate, $\alpha$, is obtained. Combining
these results with the Chudley-Elliott model, one obtains the jump
statistics, which are used to determine the transition rates,
$\bar{\Gamma}$, as the total hopping rate.

\begin{figure}
\centerline{\includegraphics[width=8cm]{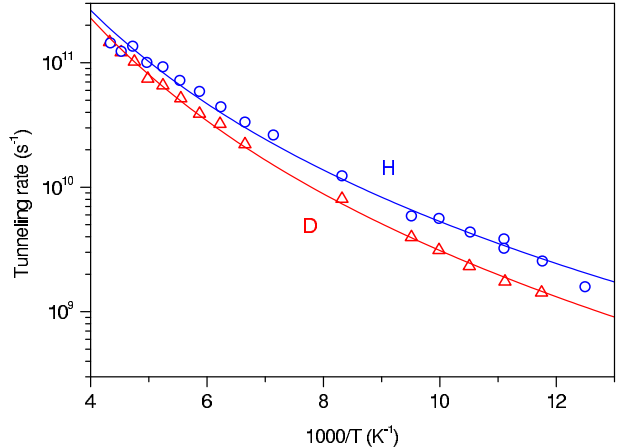}}
 \caption{\label{fig1}
  Tunneling rates for H (experiment: blue open circles; fitting: blue
  solid line) and D (experiment: red open triangles; fitting values:
  red solid line) for a coverage of 0.1~ML.}
\end{figure}

In Fig.~\ref{fig1} we show our best fit curves (solid lines) to the
experimental data (open symbols) for H (blue) and D (red). To
simplify the fitting procedure, but without loss of generality,
Eq.~(\ref{fit-function}) is recast as
\be
 \bar{\Gamma}_{\rm fit} = C \left( \frac{1000}{T} \right)^{1-2\zeta} ,
\ee
so that eventually only two fitting parameters are needed.
Accordingly, we find $C_{\rm H} = \exp(32.21)$ and $\gamma_{\rm H}
= 13.6$~ps$^{-1}$, and $C_{\rm D} = \exp(32.66)$ and $\gamma_{\rm D}
= 7.4$~ps$^{-1}$, where $\gamma$ is determined from
Eq.~(\ref{zeta}). As expected, $\gamma_{\rm D}$ is about twice
smaller than $\gamma_{\rm H}$, in agreement with the mass scaling
factor $m^{-1}$, i.e., displaying the isotopic effect. Nevertheless,
what is more remarkable is the fact these friction coefficients
imply dissipation time scales much shorter than those rendered by
the theoretical approach followed in Ref.~\cite{jardine},
based on Wolynes' dissipative transition state theory
\cite{hanggi,wolynes}. The resulting friction coefficients are
$\gamma_{\rm H} \approx 1.15$~ns$^{-1}$ and $\gamma_{\rm D} \approx
0.57$~ns$^{-1}$ (which still display the correct factor two between
them). It is thus clear that incoherent tunneling seems to be
playing an important role (as also acknowledged in
Ref.~\cite{jardine}). Nevertheless, our values for the
friction coefficients are rather close to those found for other
adsorbates considered in the literature, typically lying in the
range 0.1--5~ps$^{-1}$ \cite{jardine2}. Moreover, we have also found
that $\zeta_{\rm H} = 2.63$ and $\zeta_{\rm D} = 2.85$, which
ensures the suitability and validity of this approach to describe
the range of temperatures considered in the experiment (actually, it
should be valid even down to zero surface temperature
\cite{grabert1}).

The scenario of several loss mechanisms discussed at the end of the
previous section readily arises when considering the fact that
diffusion by tunneling is actually affected by surface coverage
\cite{graham}. In order to analyze this effect as well as to
corroborate the previous friction coefficients, an alternative
fitting is considered by using a temperature-dependent,
collisional-friction model, namely the so-called two-bath model
\cite{ruth1,ruth2,ruth3}. More specifically, this model has been
considered to accomplish two purposes: (1) to corroborate the
fitting values of the friction parameters obtained with the
Grabert-Weiss theory, and (2) to determine the effect of low
coverages on tunneling rates. This thus confers much more
reliability to the friction values found above, removing any trace
of arbitrariness.

Within the two-bath model, one bath describes the effect of surface
phonons, while the other one accounts for the collisions among the
interacting adsorbates. These two baths are assumed to be
uncorrelated and therefore the corresponding frictions are additive.
Accordingly, the total friction, now denoted by $\eta$, is a sum of
two contributions: the usual substrate friction, $\gamma$, and a
collisional friction, $\lambda$, accounting for collisions among
adsorbates ($\eta = \gamma + \lambda$). For a convenient and simple
analytical estimate of the $\lambda$ dependence on the coverage and
temperature, a simple hard-sphere model (even though we are well
aware that the dependence might be much more complex) leads to a
collisional friction given by \cite{ruth2}
\be
 \lambda = \frac{6 \rho \theta}{a^2} \ \! \sqrt{\frac{k_B T}{m}} ,
 \label{theta}
\ee
where $\rho$ is the adparticle effective radius. This equation
clearly indicates that the collisional friction is proportional to
the coverage and to the square root of the surface temperature.
Thus, this equation could also be used to extract such a coefficient
by considering the  proportionality constant a fitting parameter.
Accordingly, increasing the coverage leads to an also increase of
the collisional friction and therefore of the total friction. This
allows us to readily express our relations for tunneling quantum
diffusion as a function of coverage (with a range of validity up to
$\theta \approx 15-20\%$). In order to distinguish in the friction
coefficient the contributions coming from the surface thermal
effects and the collisions, the fitting is carried out by expressing
the exponent in Eq.~(\ref{fit-function}) as $1 - b \gamma - c\ \!
\sqrt{T/1000}$.

\begin{figure}
\centerline{\includegraphics[width=8cm]{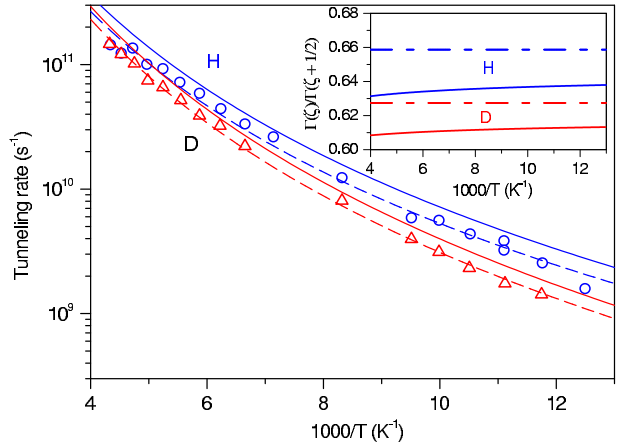}}
 \caption{\label{fig2}
  Tunneling rates for H (blue solid line) and D (red solid line) for a
  coverage of 0.01~ML; to compare with, the same curves (dashed
  curves) from Fig.~1 are plotted for $\theta = 0.1$~ML, but fitted
  according to a two-bath model (see text for details).
  In the inset, the negligible variation of the prefactor
  $\Gamma(\zeta)/\Gamma(\zeta+1/2)$ along the range of temperatures
  analyzed is also shown for $\theta = 0.1$~ML (solid line) and
  $\theta \approx 0$~ML (dash-dotted line).}
\end{figure}

Following the above best fit procedure on the experimental data, we
compute again the curves for $\theta = 0.1$~ML, finding $C_{\rm H} =
\exp(32.54)$, $b_{\rm H} = 0.388$~ps, $\gamma_{\rm H} =
13.56$~ps$^{-1}$, and $c_{\rm H} = 0.462$~K$^{-1/2}$, and $C_{\rm D}
= \exp(32.93)$, $b_{\rm D} = 0.775$~ps, $\gamma_{\rm D} =
7.34$~ps$^{-1}$, and $c_{\rm D} = 0.382$~K$^{-1/2}$. As seen in
Fig.~\ref{fig2}, the new fitting curves for $\theta = 0.1$~ML (blue
and red dashed lines for H and D, respectively) are also in
agreement with both the experiment and the previous fitting curves
(see Fig.~\ref{fig1}). Furthermore, the values of the substrate
friction $\gamma$ for $H$ and $D$ are nearly equal to those found
before assuming that all friction effects were included in $\eta$,
thus making apparent again the isotopic effect and that we are in a
moderate collisional regime. Now, according to this new functional
dependence of the tunneling rate on the coverage and surface
temperature, if we consider $c=c'\theta/0.1$, then the curves for a
lower coverage can also be readily determined. For example, in
Fig.~\ref{fig2} they are displayed for $\theta = 0.01$~ML (blue and
red solid lines for H and D, respectively).

According to this second fitting model, our preliminary assumption
that the multiplying factor $\Gamma(\zeta)/\Gamma(\zeta + 1/2)$ can
be considered to be a constant finds a justification. As seen in the
inset of Fig.~\ref{fig2}, the variation of this function along the
range of temperatures considered when $\zeta$ is expressed as a
function of temperature increases only about 1\% for H and 0.8\% for
D as $T$ decreases in the case of $\theta = 0.1$~ML (for $\theta
\approx 0$ this increment is meaningless). It is worth stressing the
fact that, as seen in Fig.~\ref{fig2}, a decrease in the total
friction (by lowering the coverage or equivalently the collisional
friction) leads to an enhancement of tunneling ($\bar{\Gamma}$
increases). This behavior is consistent with our model, since the
occurrence of tunneling should be favored by decreasing the
environmental actions on the adsorbed system. Given the high
accuracy that can be reached by $^3$He spin-echo measurements, this
is a result which would be worth checking experimentally. It would
help to understand and therefore to uniquely determine whether
tunneling play indeed a major role in the surface diffusion of light
particles, as claimed here or in Ref.~\cite{jardine} (though
approaching the problem from dissipative transition state theory).
Notice that this (predicted) behavior of the transition rate with
the coverage is opposite to previous experimental results carried
out on the same systems with quasielastic He-atom scattering and
analyzed in terms of the Arrhenius law \cite{graham}. In this case,
relatively large coverages were considered (above 15\%), out of the
range of validity of our model, where collective motion effects
could start playing a role and therefore screening any tunneling
effect. As observed in these measurements, errors increase largely
as one goes to lower temperatures and coverages, this being the
reason why possibly $^3$He spin-echo would be a better experimental
method to analyze lower coverages.


\section{Conclusion}
\label{sec4}

Summarizing, we have tackled the problem of surface diffusion of
light particles from the perspective of the Grabert-Weiss approach
to incoherent tunneling in periodic lattices
\cite{weiss,grabert1,grabert2}. More specifically, we have focused
on a series of recent experimental measurements for H and D on
Pt(111). From the feasibility of the fittings obtained to the
experimental data, we infer that the process is essentially ruled by
incoherent tunneling rather than (over-barrier) hopping. In this
regard, even though our approach differs from that considered by
Jardine {\it et al.}~\cite{jardine}, equally accurate fittings have
also been obtained for the same sets of experimental data. Indeed,
by means of this model we have determined friction coefficients,
which involve time scales several orders of magnitude smaller and
are close to the typical values found for other species
\cite{jardine2}.

In principle, the theories developed by Grabert and Weiss, and Kondo
are more general than Wolynes' one, covering the whole range of
surface temperatures involved in the experiment. Moreover, according
to Kondo, it takes into account the conduction electrons of the
surface, as it is clearly stressed in \cite{sundell}. In our
opinion, these facts could be at the origin of the discrepancies
found in the friction coefficients of $H$ and $D$ atoms on Pt(111).

In order to study the dependence of the tunneling rate with the
surface coverage and therefore the collisional friction, we have
also considered a second two-bath model \cite{ruth1,ruth2,ruth3},
valid for moderate coverages. According to this model, the tunneling
rate increases as coverage decreases. Taking into account that
$\bar{\Gamma}^{-1}$ provides us with a time scale for the diffusion
process, it means that tunneling becomes faster for lower coverages,
which would be consistent with the fact that environmental
interactions inhibit it. According to previous measurements
\cite{graham} of this effect for higher coverages than the moderate
values considered by us, the process seems to be the opposite. In
this case it could happen that collective motions are also
contributing to the transition rate, screening the effects purely
due to tunneling. Moreover, the order of the errors in these
measurements is also relatively large and therefore it is difficult
to establish a comparison. New measurements based on the more
accurate $^3$He spin-echo technique could render possibly some light
in this regard. Nonetheless, what seems to be clear is that surface
diffusion of light particles seems to be still far from a final
interpretation and description of the process. The discrepancies in
the theoretical descriptions rather suggest that more complete and
detailed simulations are needed, perhaps incorporating full ab
initio calculations with the purpose to eventually elucidate the
main mechanism involved in these processes as well as to accurately
determine the corresponding friction constants.


\section*{Acknowledgements}

This work has been partly supported by the Ministerio de
Econom{\'\i}a y Competitividad (Spain) under Projects FIS2010-22082
and FIS2011-29596-C02-01. AS would also like to thank the Ministerio
de Econom\'{\i}a y Competitividad for a ``Ram\'on y Cajal'' Research
Grant and the University College London for its kind hospitality.


\end{document}